# Public key Steganography Using Discrete Cross-Coupled Chaotic Maps

Sodeif Ahadpour
Department of Physics,
University of Mohaghegh
Ardabili, Ardabil, Iran

Mahdiyeh Majidpour
Department of Physics,
University of Mohaghegh
Ardabili, Ardabil, Iran

Yaser Sadra
Department of Physics,
University of Mohaghegh
Ardabili, Ardabil, Iran

*Abstract*— **By cross-coupling two logistic maps a novel method is proposed for the public key steganography in JPEG image. Chaotic maps entail high complexity in the used algorithm for embedding secret data in a medium. In this paper, discrete cross-coupled chaotic maps are used to specifying the location of the different parts of the secret data in the image. Modifying JPEG format during compressing and decompressing, and also using public key enhanced difficulty of the algorithm. Simulation results show that in addition to excessive capacity, this method has high robustness and resistance against hackers and can be applicable in secret communication. Also the PSNR value is high compared to the other works.**

*Keywords-component; Cross-coupled chaotic maps; Bifurcation; Statistical steganalysis; Least significant bits; Chi-square attack Introduction*

## I. INTRODUCTION

In recent years, a lot of researches have been done to

Develop image steganography. Along with the human growing dependence on computers and Internet networks, the security of information becomes more and more essential. For this purpose many methods have been developed such as, watermarking [1], cryptography [2], steganography [3]and etc. Both cryptography and steganography are two major schemes in secure information technology. However, unlike cryptography, the existence of the secret information is undetectable in steganography. This is because, the secret data is hidden in a cover file so as not to arouse an eavesdropper's suspicion [4]. Capacity, imperceptibility, security and resistance against attacks are the main criteria in steganography. In the steganography the cover-object (host file) and stego-object (the cover with secret data) created by this method must not have difference in HVS[1]. Also statistical properties of the file, before and after information embedding should have minimum differences. The cover-object for steganography could be image, video clip, text, music and etc. Due to the redundancy of data in digital images, a little change in the information cannot be detected by the naked eye, so digital images can be good covers for steganography. Several methods are used for image steganography [6]:

1- Discrete Cosine Transform, (DCT)

2- Discrete Wavelet Transform, (DWT)

3- Least Significant Bit, (LSB)

⋮

In the first two methods the pixels are transformed into coefficients, then the secret bits are embedded in these coefficients. But in the third way, secret bits are embedded directly in LSB image pixels. Those two first methods are known as transform domain and the other one is named as spatial domain. In the transform domain, major robustness is provided against changes and attacks, where spacial domain has high embedding capacity. In this paper we used discrete cross-coupled chaotic maps to embed secret data in LSBs image pixels to solve the robustness problem. In proof of our statement we will compar produced results with transform domain.

There are three types of steganographic conventions in the literature: pure steganography, secret key steganography and public key steganography. A steganographic system which doesn't rely on exchange of the secret keys are named pure steganography. This system isn't very protected in practice because its security depends entirely on its secrecy. On the other hand in secret key steganography, the sender (Alice) chooses a cover and equipped with some secret keys embeds the secret message into cover. The keys used in embedding process must be send to the recipient (Bob), so he can reverse the process and extract the secret message. However, sharing secret keys in this system introduces a new problem, i.e. the keys distribution [7]. Public key steganography reduce this problem by using two keys, a private and a public key. In this system, Alice and Bob can exchange public key by insecure channel. So both parties can determine a unique shared key, clear to both of them. The idea of public key crypto-system was expanded by Diffie and Hellman in 1976 and was published in a paper with name "New Directions in Cryptography". (Diffie, et al., 1976) [8]. Their protocol is the first system that used public key or two- key cryptography. In the Diffie-Hellman protocol a secret key is used to exchange between two parts over an insecure channel without exchanging any prior information between them. The proposed public key in this paper will be explain in the proposed method section.

Among numerous image formats, JPEG files are most popular and widely used on Internet, which is due to their small size and high quality. The abbreviation JPEG stands for the Joint Photographic Experts Group. Unlike all of the other compression methods, JPEG is'nt a single algorithm. Instead, it may be accounted as a toolkit of image compression techniques that can be altered to fit the needs of the user. JPEG may be modified to make very small, compressed images that are of relatively poor quality in aspect still suitable for many applications. Conversely, JPEG is capable of making very high-quality compressed images that are still far smaller than the original uncompressed image. JPEG is also uses primarily a lossy method for compression. So if the pixel's LSB changes during embedding procedure, our secret data will change by compressing. The proposed algorithm uses a new method to embed secret data in JPEG format also in LSB branch. Also, in many steganographic methods, this file is used as cover-image such as J-Stego [9], F5 [10], OutGuess [11]. In all of them, DCT coefficients were used for secret bits embedding. Although our method root is different from these methods, but our purpose is to demonstrate having more robustness along with capacity.

In modern steganalysis, hackers try to check statistical properties of image to realize if there are any secret bits in it. Westfeld and Pfitzman described an attack [12] based on statistical properties of image. This method is known as chi-square attack. They believed that in images LSBs are not completely random.

In one innocent image the frequencies of 2k (k is the number of image categories) pixels are not equal with the frequencies of 2k+1 pixels. If the secret bits have been equally distributed, the frequencies of POVs (pair of value) would have become equal. With this technique, they could detect the J-stego and Ezstego [13] methods. If the data embedment in LSB image pixels is based on unsystematic rule, the security of information is increased. In our work while the secret bits are spread in the whole cover-image randomly, using discrete cross-coupled chaotic maps, this method can stand against chi-square attack. Furthermore our capacity is high and the associated PSNR values indicate that robustness problem improved against hackers. The rest of the paper is orgnized as follows: In section 2 a brief description of the chaotic maps will be given. Sections 3 and 4 describe our proposed method and extraction algorithm, respectively. The experimental results are demonstrated in section 5. Finally, conclusions will be briefly given in section 6.

## II. ONE-DIMENSIONAL CHAOTIC MAPS

The general form of one-dimensional chaotic map is

$$x_n = f(x_{n-1}).$$

With these maps we can produce series $\{x_n | n = 1,2,3,...\}$ that are full chaotic and $f : I \rightarrow I$ ($I = [0,1]$). These signals are like noises but they are absolutely certain [14]. Chaotic maps have many interesting features such as sensitivity to the initial conditions and system parameter, ergodicity and mixing properties. Because of these properties, they have been used to generate random numbers. With a tiny change in their initial values the generated series are completely different. These series are deterministic, meaning that if we have the initial values, system parameter and the form of the map, we can reproduce them again. Under certain conditions these signals are random and nonperiodic. Thus, we can use the chaotic maps for selecting the image pixels for changing their LSBs. Here, we consider one of the one-parameter families of chaotic maps [5]. It is given by:

$$x_n = \frac{\alpha^2 (2x_{n-1}-1)^2}{4x_{n-1}(1-x_{n-1})+\alpha^2 (2x_{n-1}-1)^2} \quad n=0,1,2,... \quad (1)$$

where $x_{n-1}$ and $\alpha$ are the iterative value and the system parameter, respectively. To obtain random and nonperiodic numbers, it is better to restrict $\alpha$ in the $(0.5,\infty)$ range [5]. Since, in this range, the system demonstrate chaotic behavior, therefore, the generated numbers are nonperiodic.

## III. THE PROPOSED METHOD

After In our scheme, the secret message is a text. In the first step, since the ASCII 7-bit, we convert text in the English alphabet into a binary stream. However, for non English case Unicode is used instead of ASCII code in which, each character is coded by 16 bits. We take the binary stream as secret bits:

Secret bit = { $a_1, a_2, ..., a_n$ }

In the second step, we have to specify the location of embedding secret bits in the pixels. The image used for cover-object can be grayscale or true color. For a computer, one image is known as a grid of digits. A grayscale image is an $M \times N$ matrix but for a true color image it is an $M \times N \times 3$ (M and N are respectively the numbers of rows and columns and 3 is the number of main colors). Each element of the matrix corresponds to one image pixel. We use two discrete cross-coupled chaotic maps to locate image pixels whose their LSBs will be used for embedment. For this purpose, to consider a two-dimensional chaotic system which is defined as follows:

$$\begin{cases} x_n = f_1(x_{n-1}) & n=1,2,... \\ y_n = f_2(y_{n-1}) & n=1,2,... \end{cases} \quad (2)$$

that $f_1$ and $f_2$ are the one-parameter families of chaotic maps(Eq. 1). Therefore, the $\alpha_1, \alpha_2, x_0, y_0$ are our secret keys. These keys must be shared between sender and receiver as secret keys. $x_{n+1}$ and $y_{n+1}$ generate random numbers in the [0,1] interval. The output generated by the Eq.(2a) multiplied by a crus-coupling factor R is fed to the Eq.(2b) as the input (initial conditions) and vice versa (see Fig. 1). R is our public key. Because of the chaotic maps nature, none of the $[x_n, y_n]$ pairs are repeated.

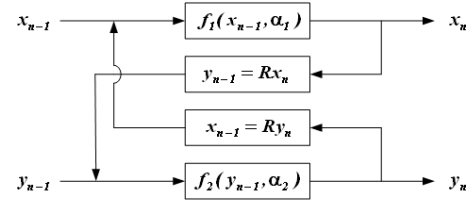

Figure 1.  The block diagram of cross-coupled one-parameter families of chaotic maps.

Next, the $\{x_n\}_{n=0}^{\infty}$ values and the $\{y_n\}_{n=0}^{\infty}$ values that are in the range [0,1], must be converted to the [1,N] and [1,M] ranges, respectively. To obtain $X_n$ and $Y_n$ must converted $x_n$ and $y_n$ to [1,N] and [1,M] using the following equations:

$$\begin{cases} X_n = [\,x_n \times N\,] + 1 \\ Y_n = [\,y_n \times M\,] + 1 \end{cases} \qquad (3)$$

Where, [ ] demonstrates the greatest integer part, $X_n$ the rows and $Y_n$ the columns of the elements. However, when converting numbers with 15-digit accuracy to finite intervals [1,M] and [1,N], the existence of duplicated numbers are inevitable. So to extract the secret message in embedded form we have to delete the duplicated numbers. After specifying elements of image we embed $a_i$ ($a_i \in \{a_1, a_2, ..., a_n\}$) in their LSBs. Now, if stego-matrix is saved in jpg format, the embedded secret message will be lose (see Fig. 2).

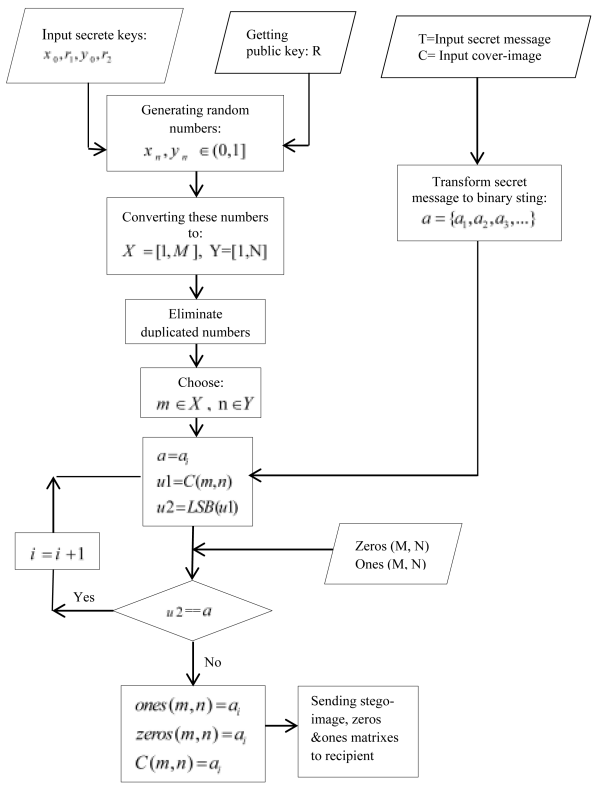

Figure 2. The block diagram of proposed method.

The last step (for solving this problem) is using one and zero matrixes in the algorithm. These two matrixes are used in the same size of cover image. After embedding the secret bits in the cover-matrix LSBs, the LSBs can change by one digit or can remain unchanged. For those pixels that changed, we write these changes in the same pixel in the two matrixes one and zero. For example, if LSB in the pixel (m,n) is zero and secret bit is $a_i = 1$, we will embed this secret bit in this pixel of cover-matrix and also in the (m,n) pixel of one and zero matrixes. In this way we will know the pixels that are changed in the stego-image. The generated zeros and ones matrixes will be sent to the recipient as secret keys.

### III.I THE DIFFIE-HELLMAN KEY EXCHANGE ALGORITHM

Used Algorithm for key exchange Alice must do the following steps:
1a) Using $\alpha_1, \alpha_2, x_0, y_0$ to compute $x_1$ and $y_1$ between (0,1).
2a) Produce $x_2 = Ry_1$ and $y_2 = Rx_1$ and so on.
3a) Calculate ones and zeros matrixes (Alic public key) and send these two matrixes to Bob.
And Bob must do the following:
1b) Choose cross-coupling factor R (Bob public key) and send to Alice.
2b) Also Bob use $\alpha_1, \alpha_2, x_0, y_0$ to compute $x_1$ and $y_1$.
3b) Get ones and zeros matrixes and calculate with stego-matrix.

The security of this algorithm is depend on the power of the discrate algorithm and the size of key space. Anyway, the Diffie-Hellman convention is considered secure against brute force attack if $\alpha_1, \alpha_2, x_0, y_0$ are chosen properly (Diffie, et al., 1976).

This method has many advantages. Maybe you say that if we have one and zero matrixes, we can extract secret message. But we note that firstly only the changed LSBs in the stego-image was marked in one and zero matrixes. Secondly with these marked {0,1} we can construct infinite permutations. Thirdly, JPEG images are compressed in saving, on the other hand, pixel LSB magnitude changes one unit in maximum, but its JPEG compression magnitude doesn't change by one unite. Means that without zero and one matrixes, rusher cannot decide if the extracted information is meaningful or just part of the natural randomness.

### IV. EXTRACTION PROCESS

In order to extract the secret message, one needs to implement the reverse algorithm of the embedment. If we have the initial values as secret keys, R as a public key, and know the form of functions, we can obtain the produced series in embedding algorithm. After finding the location of embedment, we have to compare ones and zeros matrixes. Whenever the pixel of ones matrix is as the same as zeros matrix pixel's, we realize that the LSB in the same pixel of the stego-matrix was changed, and if the pixels of ones matrix and zeros matrix are different, it is clear that this pixel in stego-matrix didn't change during embedding.

Suppose we identified the pixel (m,n) from stego-image that secret bit must be in its LSB, comparing two corresponding elements from ones and zeros matrixes help us to know if the LSB was changed in this pixel at the embedding time or not. In other word if the (m,n) element in the ones and zeros matrixes is same, the corresponding pixel from stego-image was changed. Fig. 3 shows this fact for $5\times5$ matrixes, in the Now the receiver know which LSB in which pixel are the secret bit. If we integrate these LSBs, we can discover the characters in the used form.

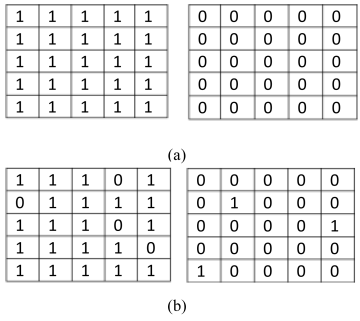

(a)

(b)

Figure 3. (a) Ones and zeros matrixes, we see that all of the elements are totally different.(b) Now look at the elements {(1,4), (2,1), (2,2), (3,4), (3,5), (4,5)} from these two matrixes, they are the same, i.e. in the stego- matrix these elements were changed, so the embedded secret bit in these pixels must be picked from ones or zeros matrixes.

## V. EXPERIMENTAL RESULTS

In this section, we demonstrate the performance of our proposed method and compare it with that of F5 , J-Stego. For this purpose, five standard 512×512 grayscale images with different textural properties were taken as cover-images (see Fig. 4).

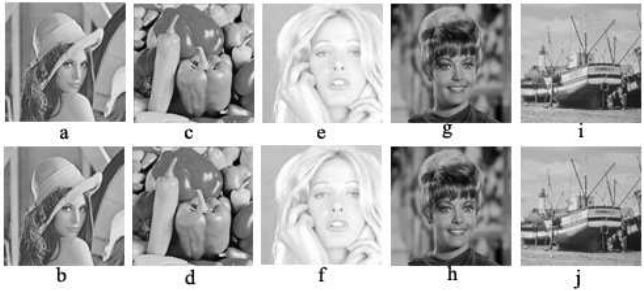

Figure 4. Upper part, from left to right: (a) Lena, (c) Peppers, (e) Tiffany, (g) Zelda, (i) Boat, are the cover-images. Bottom part, from left to right: (b) Lena, (d) Peppers, (f) Tiffany, (h) Zelda, (j) Boat, are the stego-images.

### V.I CAPACITY

The capacity is the size of the data in a cover image that can be modified without deteriorating the integrity of the cover image. The steganographic embedding operation needs to preserve the statistical properties of the cover image in addition to its perceptual quality. In the proposed method, one LSB from each pixels has been used. The number of pixels in one 512×512 image, is greater than 260000 pixels, on the other hand in using one LSB from each pixel to embedding, in maximum state only half of the desired pixels change with respect to the secret bits. Therefor, if we insert as many secret bits as half of the total pixel numbers, only a quarter of LSB image pixels will change (see Table1).

TABLE I. COMPARISON OF CAPACITY(BITS) FOR VARIOUS EMBEDDING\NEWLINE ALGORITHMS

| Test Images | Our Method | J-Stego | F5 |
|---|---|---|---|
| Lena | 47600 | 32998 | 33026 |
| Peppers | 52448 | 34295 | 34074 |
| Tiffany | 52800 | 31674 | 31516 |
| Zelda | 57200 | 27557 | 27630 |
| Boat | 42912 | 38374 | 38506 |

Also, the capacity regarding the obtained PSNR value is represented by bits per pixel (bpp) and the Hiding Capacity (HC) in terms of percentage (see Table2). According to the experimental results, we can conclude that the capacity obtained by our approach is great.

TABLE II. COMPARISON OF PSNR (dB) OF THE STEGO-IMAGES CREATED BY VARIOUS EMBEDDING ALGORITHMS

| Test Images | HC (bpp) | Our Method | J-Stego | F5 |
|---|---|---|---|---|
| Lena | 0.25 | 61.90 | 33.36 | 36.94 |
| Peppers | 0.25 | 61.45 | 35.45 | 35.86 |
| Tiffany | 0.25 | 62.41 | 35.93 | 36.36 |
| Zelda | 0.25 | 61.91 | 38.31 | 38.82 |
| Boat | 0.25 | 62.42 | 35.67 | 36.23 |

### V.II IMPERCEPTIBILITY

In modern steganography the difference of stego-image and cover-image are not distinguishable even by use of advanced computers. After producing stego-image, we have to evaluate the steganography quality or compare stego-image with cover-image. This work is done by computing PSNR (peak signal to noise ratio):

$$PSNR = 10 \log_{10} \left[ \frac{C_{max}^2}{\frac{1}{M \times N} \sum_{i=1}^{n} \sum_{j=1}^{m} (C_{i,j} - S_{i,j})^2} \right] \quad (4)$$

In this equation $C_{max}$ is the maximum value of the cover-image. For example:

$$C_{max} \leq \begin{cases} 1 & \text{in double precision intensity images} \\ 255 & \text{8 bit unsigned integer intensity image} \end{cases}$$

M and N are the number of rows and columns. $C_{i,j}$ and $S_{i,j}$, are the values of the gray levels of cover-image and stego-image, respectively. If the difference of the two matrices tends to zero, the PSNR leads to infinity. Large values of this quantity implies more similarity between stego-image and cover-image. In Table 2, the PSNR value for our work and two other works are given. As can be seen, the obtained PSNR for the proposed algorithm is higher than other works by a remarkable difference.

## V.III INFORMATION ENTROPY

that can be used in characterizing the input image texture. It is given by:

$$H = -\sum_{i=0}^{255} p_i \log_2(p_i)$$

Where, $p_i$ is the probability that the difference between two neighbor pixels equals i. In steganography, the created cover-image and stego-image must have close entropy. The calculated entropies for two files are presented in Table 3. For all images, the obtained values are very close to each other showing that the added noise is very low.

TABLE III. COMPARISON OF COVER-IMAGE AND STEGO-IMAGE ENTROPY

| Test Images | Cover-Image | Stego-Image |
|---|---|---|
| Lena | 7.3938 | 7.3940 |
| Peppers | 7.5918 | 7.5919 |
| Tiffany | 6.5454 | 6.5495 |
| Zelda | 7.2668 | 7.2667 |
| Boat | 7.1237 | 7.1239 |

## V.IV ROBUSTNESS AGAINST ATTACKS

The chi-square attack is based on POV statistical analysis. Its idea is to obtain the theoretically expected frequency distribution and compare it with the frequency distribution of suspicious image. In one innocent image the LSB values are not completely random. But in one image with embedded information, the 2k pixels frequencies are equal (nearly) to the frequencies of (2k+1) pixels. We drew the probability of the embedding based on the percentages of image pixels. As shown in Fig. 5, this probability is zero for our work. But the results show that the probability of embedding data by J-Stego and Ezstego algorithms are not zero [19].

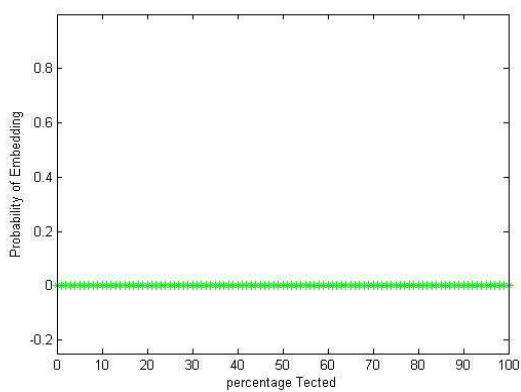

Figure 5. The detection results of the chi-square attack to the stego-image created by our algorithm for Zelda's image.

## VI. DISCUSSION AND CONCLUSIONS

The founded elements via cross-coupled chaotic maps are highly random, therefore the secret message is spread all over the image. More importantly, modifying JPEG format during compressing and decompressing, and also using ones and zeros matrixes protect our stego- file from attackers. In LSB method, when we use only one LSB for embedding the secret message, only half of the desired pixels change with respect to the secret bits. So the capacity is great for the method. Additionally, the change of color intensity is very little, hence HVS will not be able to distinguish the cover-image from stego-image. The obtained results insures the work's robustness and high secrecy, compared with other works. The obtained high PSNR value, close entropy values, for cover and stego images, and the zero probability for embedding secret information, prove the introduced method to answer the needs of today's secure communications.